# Simulations of a silicon pixel based on MOS Deep Trapping Gate Principle

**Nicolas T. Fourches**[a*] **and Wilfried Vervisch**[b]

[a] *CEA Saclay, IRFU ,SEDI
91191,Gif/Yvette,France ,
E-mail: nicolas.fourches@cea.fr*

[b] *Aix-Marseille Université, CNRS, IM2NP, UMR 7334
13397, Marseille, France*

*E-mail*: (corresponding author): nicolas.fourches@cea.fr

**ABSTRACT:** The concept of the deep trapping gate device was introduced fairly recently on the basis of technological and transport simulations currently used in the field of classical electron devices. The concept of a buried gate containing localized deep level centers for holes (Deep Trapping Gate or DTG) renders possible the operation of this field effect pixel detector. One alternative to Deep Level introduction is the use of a quantum box, which is a hole quantum-well and an electron barrier. In all of these cases the buried gate modulates the drain-source current. This principle was formerly evaluated with realistic simulations parameters and this shows that a measurable signal is obtained for an energy deposition of a minimum-ionizing particle within a limited silicon thickness. In this work a quantitative study of the response of such a pixel to Minimum Ionizing Particles. The influence of some parameters such as the thickness of the pixel and its lateral dimensions, on the operation of the pixel is studied here using current available simulation tools, in quantum mode when a narrow Ge layer is used as a buried gate. A bias sequence is introduced here to separate the operation of the pixel in detection and readout mode. We will study here the potential techniques usable for the fabrication of the device with and try to find the most relevant for this device. The first technique introduced is deep impurity implantation, followed by annealing. Recent work on quantum boxes and dots opens the possibility to the use Ge boxes as a DTG. The DTG could be fabricated with ion implantation and alternative methods such an epitaxial growth DTG can be reasonably considered. If the related bottlenecks are overcome the pixel should be a good candidate for high luminosity particle collider inner detectors, provides that its potential radiation hardness is proven sufficient for these applications.

# Contents

## 1. Introduction and description of the structure

The next generation of particle physics experiments will require high precision inner vertex detectors with very good radiation hardness. Recent relevant silicon pixel developments include the CMOS[1] sensors [1] the DEPFET[2] [2] as well as the classic HPD[3]. The first two solutions may lead to a good spatial resolution although DEPFETs require a high voltage process and CMOS sensors are very sensitive to bulk damage [3]. This excludes them from some applications such as LHC[4] upgrades or Super-LHC. To make further progress a new device was recently proposed [4][5]with the goal of decreasing pixel size and improved radiation tolerance the Deep Trapping Gate (DTG) principle. Reduced dimensions are necessary to improve point-to-point resolution and also to limit Hall Effect [6] due to the magnetic field inside the detector. The main feature of the new pixel detector is to replace the charge collecting diode by a new kind of buried gate, able to selectively trap charge carriers, named here a DTG. Two alternative structures have been proposed so far. The two have the same purpose. They localize holes in a buried gate located under the MOS n-channel, which controls the Source-Drain current in similar way as the upper gate (Fig.1). The operation of the structure has been checked for thick (10 μm) and thin structures (1 μm). If one seeks a detection efficiency close to one (99%) (for Ge see [7] ) Geant 4 simulations for Minimum Ionizing Particles) thick structures are necessary. Thin structures can be used too at the expense of detection efficiency, as they remain sufficiently sensitive in terms of Signal/Noise ratio (see CMOS MAPS results). In a thick structure the detecting volume is the device substrate that should be slightly doped p-type material (net doping concentration below $10^{14}/cm^3$,) an n-well can be added at the bottom of the structure using a deep implant or a layer deposited on the flipped structure.

---

[1] complementary oxide semiconductor
[2] depleted p-channel field effect transistor
[3] hybrid pixel detectors
[4] large hadron collider

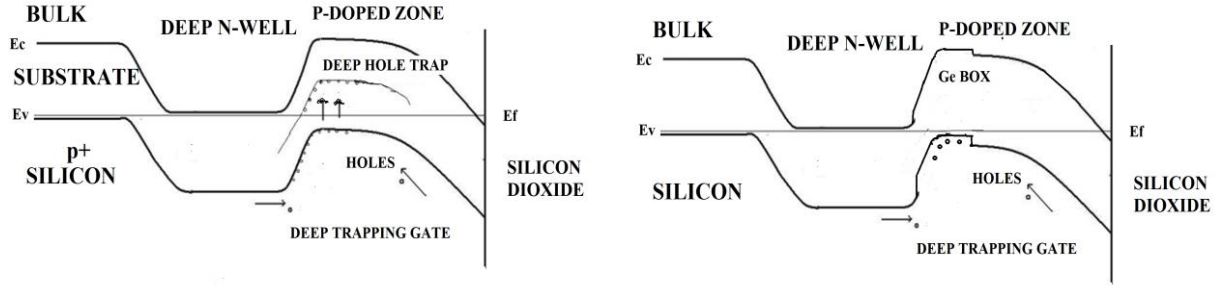

Fig. 1: Transverse view of the structures: left: band diagram of the trapping MOS structure showing the deep trapping gate. The concentration of hole traps is of the order of $10^{18}/cm^3$. The hole traps can be replaced by a quantum well (right). The potential barrier makes the holes migrate towards the DTG and electrons escape towards the bulk and channel, leaving the DTG positively charged. In this case the doping level in the n-well (bulk) is $10^{16}/cm^3$ and the p-type substrate is of the order of $10^{15}/cm^3$, the thickness of the structure is of the order of 1 µm.

In a thin structure the presence of a deep-n-well close enough to the MOS channel introduces a potential barrier large enough to separate the electron-hole pairs without the need to apply an external field. Preliminary process simulations based on ATHENA (Silvaco) have assessed the validity of medium energy ion implantation to create the deep-n-well inner layer.

**1.1 Deep level based structure**

The first structure uses the properties of deep-hole traps located in the buried gate (Fig.1 left). The way that is studied for this purpose is based on ion implantation combined by annealing. The impurity used for this study is substitutional zinc, which give rise to hole traps within the silicon bandgap ($E_v+0.27$ eV, $E_v+0.6$ eV) [8], The migrating holes get trapped at these centers and modify the charge state of the buried gate (named here Deep Trapping Gate). A preliminary study of this deep gate was published in[4]. The emission rates of these centers are low enough at 300 K to induce a significant memory effect. As a matter of fact for a hole capture cross section of $10^{-14}$ $cm^2$, an effective DOS (effective density of state in the valence band) of the order of $10^{19}$ $cm^{-3}$, a thermal carrier velocity of $10^6$ $cm.s^{-1}$ at room temperature 300 K, the emission rate for a 0.27 eV level is of the order of one microsecond and for a 0.60 eV in the hundreds of ms range. This proves that the simulations results previously obtained are realistic if the close to midgap hole trap is present in sufficient concentration. The device proved functional with peak Zn concentration of $10^{18}$ $cm^{-3}$-$10^{19}$ $cm^{-3}$, which is just slightly lower than the effective DOS in the valence band. A thorough experimental study of these implanted layers will be necessary to work out a reliable fabrication procedure.

**1.2 Quantum Box based structure**

Fig. 1 right is the corresponding band-diagram for a QBTG (Quantum Box Trapping Gate) (for further studies see [9][10] with a germanium-enriched box ($Si_{1-x}Ge_x$, x~1) or Ge only for a thin layer grown on the silicon substrate. The critical thickness being of the order of 2nm for a germanium only stable layer one should expect a relaxed layer with a thickness of 20 nm. It is not straightforward to determine the VB (Valence Band) and CB (Conduction Band) offsets in these particular cases. In practice, the use of *PECVD* for Ge and Si deposition should be one of the preferred methods of engineering the bandgaps and band offsets to fit our needs. Using the graphs of ref for a relaxed Ge layer grown on Si the VB offset is higher than 0.2 eV acting as a well and there should exist a barrier for the CB of the order of 0.12 eV. The calculation of the valence states is more difficult because of the VB splitting when the layer is compressed. Figure 19 in reference [10] shows that ΔVB can vary from 0.24 eV if a relaxed Ge layer is grown on strained Si to 0.74 eV when relaxed Si is grown on a strained Ge. For each case a positive ΔVC exists in the Ge layer so that the Ge layer produces a BC barrier for electrons (Table 1).

Table 1: short summary of the literature data on Ge on Si, from [10] and references herein], Δ is the band offset.

|  | Relaxed Si layer | Strained Si layer (substrate) |
|---|---|---|
| Strained Ge layer | ΔVB=>-0.66-0.74 eV Quantum Well | ///// |
| Strained Ge layer | ΔVC=+0.25 eV     Quantum Barrier | ///// |
| Strained Ge layer | Eg=0.68 eV | ////// |
| Relaxed Ge layer | bulk properties | ΔVB=-0.24 eV Quantum Well |
| Relaxed Ge layer |  | ΔVC=+0.6eV Quantum Barrier |

Further on in this work we have chosen to simulate the structure based on the Ge quantum box DTG, as it is the most straightforward way to check the validity of this alternative to the deep level version.

## 2. Operation of the structure

For the two thin structures a deep n well is necessary to build up the potential and the necessary electric field that will separate and drift and the generated carriers. In the space charge zone the electrons drift downwards to the substrate and holes towards the channel. For this reason such a short structure can be operated in a steady state bias: source grounded through a resistor or current source and drain and gate set at a positive voltage. This is not the scheme used for the thick structures, where the adequate field is applied trough the device, using the substrate, the source the drain and the gate.

Fig. 2 summarizes the operation of the TRAMOS for the case of a relatively thick structure (> 1 μm). In the detection mode electron-hole pairs are generated in the bulk and in the readout mode the channel current act as a probe for ionizing particle interacting in the fiducial volume (detection volume).

This new pixel configuration can be compared with a standard CMOS pixel (the simplest MAPS architecture, Monolithic Active Pixel Sensor) using a 3T configuration. Fig. 3 explains the way a 3T MAPS can evolve into a 1T TRAMOS Cell. The reduction in the number of transistors allows this pixel to shrink in area along with the scaling rules of current CMOS technologies. A preliminary remark should be made about device simulation. These were made with a 2D simulation code. Earlier work shows that these simulations can be extrapolated to a 3D device if some attention is made to the physical quantities involved. For example the signal formation due to a Minimum Ionizing Particle (80 e-h pairs per micron in Si) has been studied on a PIN structure (for early work see [3]), allowing that a "calibrated" code is used for these simulations :(by calibration one should understand that the physical quantities simulated have a value which fits to these of an identical physical device measured or calculated). A similar method is used to check the validity of the current and voltage through the MOS structure, which in the 2D simulations is a 1μm equivalent large device.

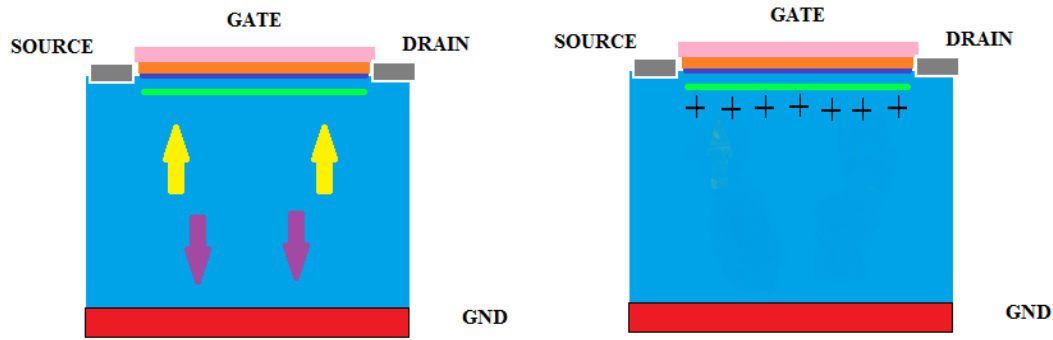

Fig. 2: Operation of the TRAMOS pixel. On the right the upper gate is negatively biased with 0 bias on the source drain and substrate. The holes generated by the particle track in the active volume (blue) drift to the upper side of the structure (yellow arrows) and become localized or trapped in the (Ge or SiGe) lower gate (green), the electrons (purple arrows) escape towards the substrate (grounded). In this detection mode the drain and source are grounded. In readout mode (left) the source is connected to a grounded load and the drain and gate are connected to Vdd (+3.3 V). The net positive charge contained in the lower gate modulates the source-drain current allowing the detection of the previously ionizing particle track or hit. The thickness of the device can vary from 1 μm to 10 μm.

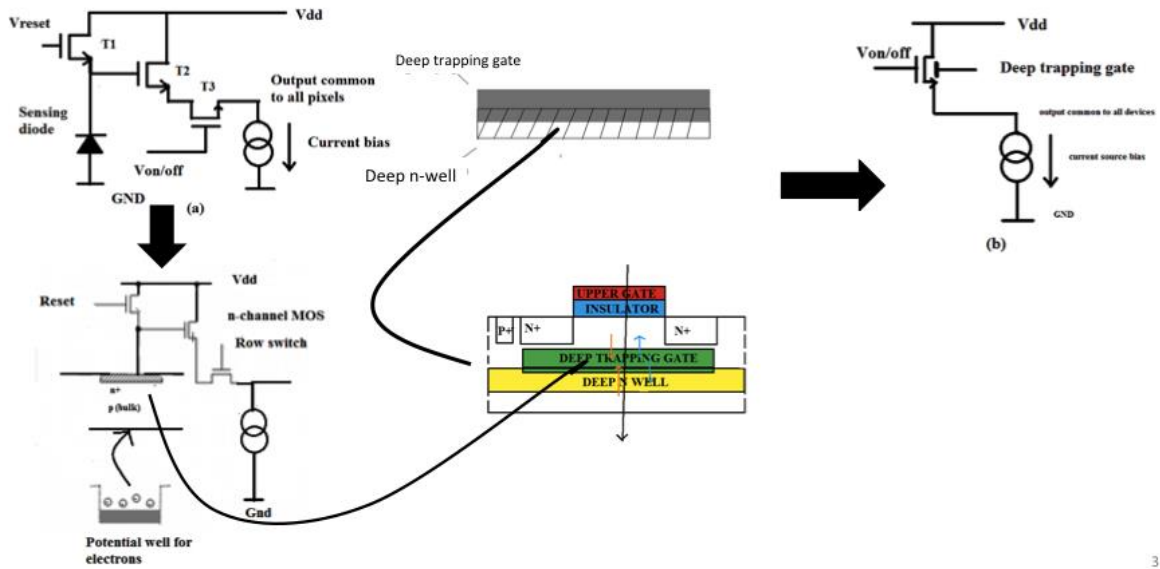

Fig. 3: evolution of the 3T CMOS sensor too the 1T TRAMOS cell. In this case the thin structure is represented. The readout can be made in the same way as it can be done with a 3T cmos pixel.

## 2.1 Quantitative study

A quantitative method will validate the principle of the device. We have already published some simulation results in [4] for the trapping version and so, as the following study is valid for the two versions, we focus on Ge layer DTG structure. In the 3T CMOS pixel configuration the charge is shared between the gate of the amplifying transistor, the n+ electrode and the source of the reset transistor. This results in a Conversion Factor of the order of 50-100 μV/e when no major additional amplifying is implemented [11]. In the case of the TRAMOS the determination of the conversion factor can be done by simulation or analytically. However a valid analytical calculation is difficult to obtain, as the device operates as an inversion mode MOS device for the upper gate and as a depletion-enhancement Field Effect Transistor for the buried gate. On the other hand the values of the capacitances can be obtained relatively easily for the upper-gate/channel/buried gate structure. In Table 2 the values of the simulated conversion factors are given w.r.t. MOS gate length at a constant gate width of 1 µm. The Silvaco package was used here to determine parameters such as the Buried Gate transconductance. The buried gate potential variation was estimated using an 80 e charge deposited. It can be concluded from this Table that a CVFs' close to the ones obtained in CMOS sensors are possible with submicron MOSFETs (=< 0.25 µm) which is obviously within the reach of actual technologies.

The steady state current when the device is biased is of the order of 0.3 mA for the 0.1 µm gate length device. With a bandwidth of 100 MHz, Johnson noise from the source resistor or Schottky noise from the channel current are of the order of $1.82 \times 10^{-8}$ A and $10 \times 10^{-8}$ A respectively, who give a 600 µm at the source (< 1 mA rms). For a MIP and a thin 1µm thick detector the Signal to Noise Ratio is of the order of 10 for a gate length of 0.1 µm.

Table 2: characteristics of the TRAMOS pixel with a 1µm channel width downscaled from a 1 µm channel length to a 0.1 µm. The steady state transconductance of the buried gate was determined using DC simulations with the Ge layer acting as a control gate. The Buried Gate capacitance/Upper Gate was determined with sufficient accuracy. During the readout or detection the upper gate is maintained at a fixed voltage (+3.3 V during readout and -3.3 V during detection. The simulations were made with a Silvaco™ package. The value of the source resistance was set in order to obtain a adequate CVF with limited source current during device simulations and improve conversion, which is more difficult with an ideal fixed current bias.

| L (µm) Gate Length | Buried Gate transconductance (µA/V) | Buried Gate-Upper Gate capacitance | ΔV (buried gate) for 80 e charge deposited | ΔV (source) for 80 e charge deposited and a 5 kohms source resistance | Conversion Factor in µV/e (source of the transistor) 5 kohms source resistance |
|---|---|---|---|---|---|
| 1 µm | 52 | 5.31 fF | 3.01 mV | 785 µV | 9 µV/e |
| 0.5 µm | 60 | 2.65 fF | 6.00 mV | 1.56 mV | 19.5 µV/e |
| 0.25 µm | 55 | 1.32 fF | 12.0 mV | 3.3 mV | 41.25 µV/e |
| 0.1 µm | 60 | 0.531 fF | 24.1 mV | 7.23 mV | 90.4 µV/e |
| Gate width 1µm | Simulated using Silvaco ATLAS | Gate width 1µm | Silvaco ATLAS | Silvaco ATLAS simulations | Silvaco ATLAS, Gate width 1µm simulations |

In Fig. 4 the thick structure used for simulation with a 100 nm gate length shows that the germanium gate has a volume of approximately 100nm x 20 nm x 1000 nm = $2 \times 10^{-15}$ cm$^{-3}$. This imposes a doping level inside the buried gate higher than this value. In these simulations, a p-type boron doping of $2 \times 10^{17}$ cm$^{-3}$ is used. In the detection volume a low net doping level is required to increase depletion at the voltages used in the device. Practically a $10^{14}$ cm$^{-3}$ doping level has been set for the simulations. The volume of this area is approximately: $10^{-3}$ cm x $10^{-5}$ cm x $10^{-4}$ cm = $10^{-12}$ cm$^3$, which means that at this scale only a few hundreds of doping atoms are present in the volume. This will mean that at this scale almost full depletion is attained. On the upper side of the structure, the maximum electrical field in the oxide is of the order of: $3.3V/5\times10^{-7}$ V/cm equal to 5 MV/cm <10 MV/cm which is sufficiently low to avoid breakdown (ref [12]). The simulation of the structure confirms this result with a field of 3-4 MV/cm in the gate oxide (Vgate=+/-3.3 V). Specific work in this field can be found in references [12][13][14] and technologies in [15]. The simulated structure is compatible with the characteristics of available technologies.

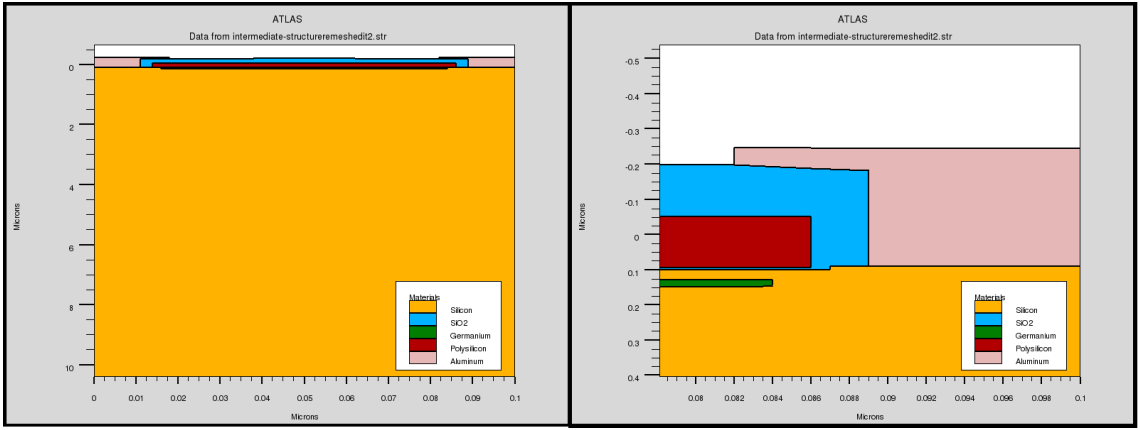

Fig. 4: Diagrams of the structures used in the following simulations. On the right the deep germanium gate of thickness 20 nm under the MOS channel. The volume for the detection of a charged particle is 10 μm deep under the germanium gate and 0.1 μm and 1 μm wide (left). In this case the deep n-well is located on the bottom of the device (10 μm to 10.5 μm) approximately and grounded for all simulations. The gate oxide is set to a value approximately equal to 5 nm.

### 2.1.1 Simulation conditions: quantum mode

Preliminary work has been made for PIN structures using a sandwich Germanium layer. When reversibly biased electron-hole pairs generated during a first time duration results, using quantum simulations in a hole current transient through the structure, which shows that the Ge layer acts as a retention layer for the holes, in fact it just act as a trapping centre. The quantum simulations can be done with a Schrödinger-Poisson approach or by using a Quantum Density Gradient Model [16], the latter being very close to a Drift Diffusion Model, with a correction potential. For hole transport: $J_p(\text{vector}) = qD_p(\nabla p) - qp\mu_p \nabla(\Phi - \Lambda) - \mu_p p kT \nabla(\ln(p_i))$ [1]

$\Lambda$ is defined by: $\Lambda = (-\gamma \hbar^2/12m)(\nabla^2(\log(p)) + 1/2(\nabla(\log(p)))^2)$ [2]

$\gamma$ is a fit factor and $\hbar$ the reduced planck constant in these equations.
The above relations show that the transport equations are close to the Drift-Diffusion Model with a corrective potential $\Lambda$ and $\Phi$ being the electrostatic potential. The simulations of the appropriate structures have been done with some parameters of the initial material set to take into consideration possible degradation due to defects. A 100 ps lifetime for the electrons and holes has been used throughout the final study. With this value and a carrier velocity of $10^7 \text{cm.s}^{-1}$ at room temperature carriers can move up to 10 μm, a distance that is sufficient to retain a reasonable charge collection by the buried gate.

### 2.1.2 Readout and Detection: bias set

In Fig. 5 the band diagram of the structure is given for the detection mode bias set. A plus 3.3 Voltage is applied on the upper gate while the substrate, source are grounded and the drain connected to the +3.3 rail. The band diagram corresponds to the middle of the device from source to drain. As it can be seen on the left figure the thin germanium layer acts as a potential-well for holes and is a barrier for electrons (CB in black). A value of 8 eV obtained with the device simulation corresponds to the silicon dioxide covering the upper polysilicon gate which is this case is metallic (CB=VB). As the gate oxide is to thin, it can be represented in these figures. The holed in the VB should move towards the substrate in this case. One can easily conclude that when the detection mode is on (-3.3 Volts on the upper gate and the rest of the electrodes grounded) the hole will move towards the channel (on the left of the device).

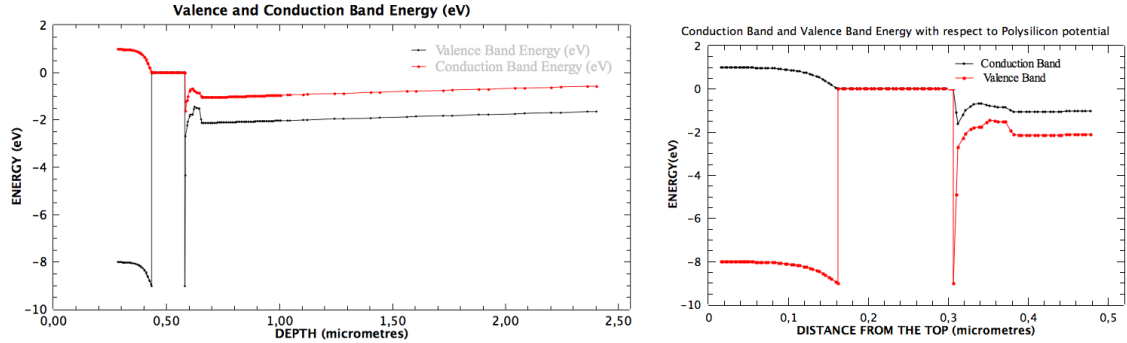

Fig. 5: Energies of the valence band and conduction band ends of the simulated structure. On the left is represented the first 2.5 μm from the top of the 10μm structure. On the right a detailed profile show the valence band well induced by the presence of the germanium layer. The TRAMOS structure is on its readout mode (positive bias at the top of the structure).

In this mode the well for holes made by the germanium layer is sufficiently deep so that the retention of these carriers is possible with the bias applied on the structure. Note that a source to drain electron-current flows in the channel delimited by the Ge layer and the Si/SiO$_2$ interface with the gate-oxide in this mode. Because of CB barrier there is no current flow in the germanium layer.

## 2.2 Response to a hole-electron generating pulse

We have made a set of simulations using a C-written routine inserted in the Silvaco script that simulates the ionization of so called Minimum Ionizing Particle. To increase the detection efficiency (probability of obtaining a hit in a pixel) up to more than 98 % one should have an active pixel thickness above 10 μm [12]. So a signal of 800 e has been used in the simulations. Figure 6 show the bias scheme (left) and the electrical response of the device (right), the lifetime of the carriers is set to 1ns. The simulated hit occurs when the time is at about 1.5 ms (see Figure 6 left). The rest of the time the device is set in readout mode. In the real world this would be the opposite with detection duration much longer than the readout time. For this configuration the signal magnitude is 90 mV at the source (source follower).

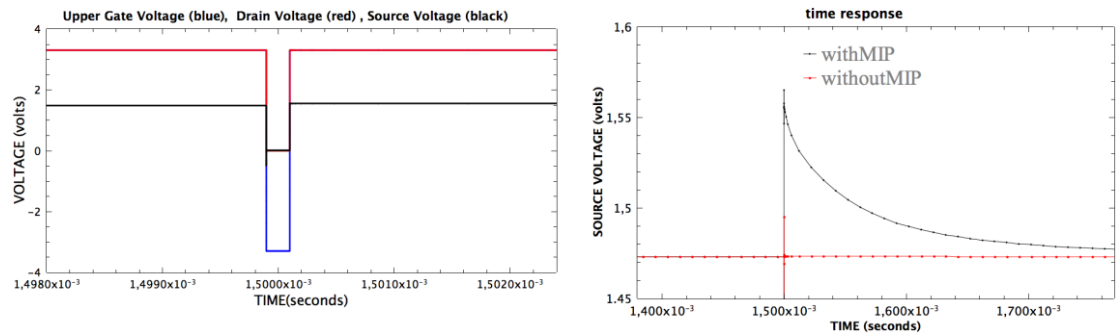

Fig. 6: Simulated device response to a MIP (minimum ionizing particles, generating 800 e/h pairs in the 10 μm thick detecting layer. The source resistance is set to 5 kohms and the generating step is of 4ns duration, in detection mode (Vs=Vd=Vsub=0 V, Vg=-3.3 V). In readout mode (see Figure 5), Vsub=0 V, Vd=Vg=+3.3 V, Vs=1.47 V steady state). The chronogram of the bias scheme is represented on the left hand side. The pixel is set from readout mode, then detection mode for 200 ns and then in readout mode. The source output signal is only visible in the right hand figure. The dimensions of the MOSFET is W/L = 1μm/0.1μm.

The device is fast enough to detect signals that have 4ns duration. Then the readout can be made many microseconds after the hit due. There is a natural reset in this case, which lasts in this case a few hundreds of microseconds. A fast reset scheme was introduced earlier for the Deep Trap version but the some efforts are still going on to simplify this. The use of the natural reset in the collider experiments can be possible if the occupancy remains sufficiently low. Here there should be no multiple hits within one hundred microseconds, which should be possible for a submicron-meter pixel. In collider (LHC) experiments the hits rate for the inner detector (CMS) is of the order of 2 GHz/cm$^2$. For a 1µm$^2$ pixel this corresponds to a hit rate of 20 hits per second. This shows that a "natural reset" is a valid option.

The signal magnitude of 80-90 mV for 800 e-h pairs give a CVF of 100-122.5 µV/e which is close to the value calculated in Table 1 for a pixel of same dimensions 90.4 µV/e). This validates this pixel design.

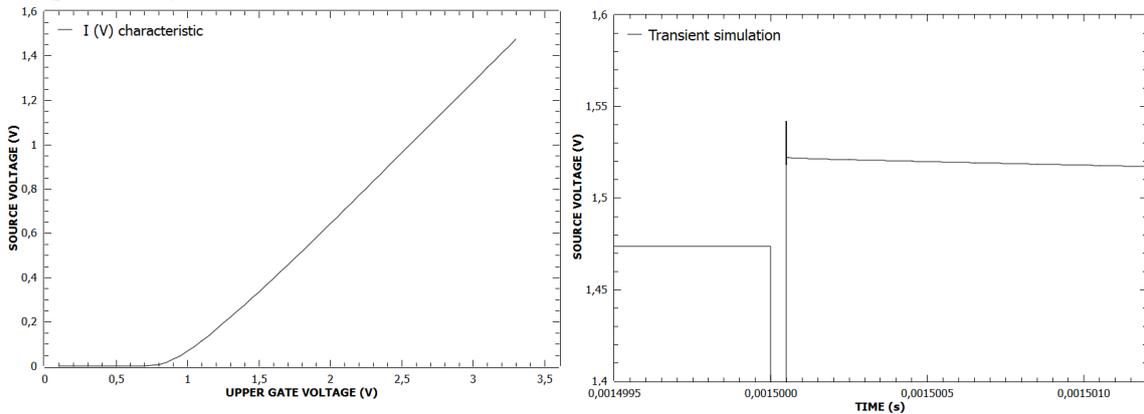

Fig. 7: Source voltage versus upper-gate voltage characteristic of the device with 100 ps lifetime materials on the left. On the right the source voltage (source follower with a 5 kohms load) is represented versus time for an equivalent MIP interaction during the detection time (50 ns in detection mode). The magnitude of the response exceeds 50 mV and a strong memory effect is visible well above 1 µs.

Other simulations studies show (Fig.7) that with 100 ps lifetime the signal is only slightly lower of the order of 60 mV. The minimum detection time is set to the order of 50 ns. This makes the device a fast detection device that could fulfil the constraints of the high luminosity colliders. It also can operate with low lifetime materials, this show that the pixel has a good hardening potential for bulk radiation induced defects.

## 3. Fabrication issues: ion implantation and material growth methods

For both DTG concepts the expected bottlenecks are related to material engineering needed for manufacturing the device. They both need a deep-n-well for adequate operation. Two main fabrication solutions may be proposed. Electrical characterization of the material/device is necessary at this step together with structural and optical properties evaluation. Note that the band offsets (VB and CB) energies in the case of a silicon-germanium-silicon structure depend on the growth conditions [10] and should be experimentally determined.

If we consider the buried deep n-well,, necessary for the correct operation of the thinnest device, it can obtain by P or As implantation at high energy (> 200 keV) into the silicon substrate followed by a thermal annealing as in a standard CMOS processes. For a 10 µm deep n-well the ions implantation method is not the best as the ions energies in silicon should then be of 40 MeV (P and As) as SRIM simulations indicate (Fig.8). This energy can only be attained in accelerator based implanters that are not in current use in the semiconductor industry. It then more reasonable to used n+ epitaxial silicon on a silicon substrate followed by another p-type silicon layer of 10 µm in thickness, which is a standard microelectronic fabrication process.

The fabrication of the quantum box is different as the Ge dose should be above $3 \times 10^{17} cm^{-2}$ (a Ge peak concentration above $5 \times 10^{22}$ cm$^{-3}$) to obtain a Ge concentration close to the silicon concentration. At these doses SRIM (ions + recoils) simulations show that a peak energy of 3600 eV per silicon atom is deposited in the target silicon material. That means that the corresponding region becomes amorphous according to the Critical Damage Energy Density model (CDED [18] [17] for Ge [18] for Si). This contrasts with epitaxial layer deposition methods from which the electronic properties of the Si/SiGe/Si structure provided until now were derived (see first paragraph). Further studies are being planned to avoid amorphization such as an appropriate annealing during or after implantation, this will constitute one of the first bottleneck to be overcome. The electrical and electronical properties will also be studied in order to get information on the evolution of the direct bandgap of the Ge or SiGe layer. If this fabrication process is not effective the alternative will then be to use epitaxial layers instead.

The introduction of the impurity for and implanted DTG (Deep Trapping Gate) is more difficult. As a deep impurity, some studies [8] indicate, for instance that Zn induces two deep levels in the bandgap of silicon, one at $E_v + 0.60$ eV, the other at $E_v + 0.27$ eV. These two levels act as hole traps. There are two requisites. First the Zn ions should be in substitutional sites and second that the peak concentration reaches more than $10^{18}$ cm$^{-3}$-$10^{19}$ cm$^{-3}$ with a sharp distribution. Although these conditions seems to be very stringent SRIM and SUPREM, ATHENA (Silvaco) shows that such as a impurity this is possible if some post implantation thermal annealing is made, the unknown figure being that ratio of deep electrically active to the total impurity. This will be at the expense of a thermal budget. If this implantation technique is not effective, then the alternative should be to growth a thin highly Zn doped epi-layer, with a low following thermal budget. Table 3 summarizes the implantation process at low energies.

Table 3: concentrations, introduction rates for Zn, Ge, P ions implanted in silicon deduced from SRIM simulations.

| Ion | Energy (keV) | Introduction Rate (cm$^{-1}$) (approximate) | Dose (cm$^{-2}$) | Concentration (cm$^{-3}$) |
|---|---|---|---|---|
| Zn | 100 | $4 \times 10^4$ | $5 \times 10^{13}$ | $5 \times 10^{18}$ |
| Ge | 100 | /// | $3 \times 10^{17}$ | $5 \times 10^{22}$ |
| P | 300 | $4 \times 10^4$ | $5 \times 10^{12}$ | $2 \times 10^{17}$ |

For the deep n-well the P or As dose is lower than the critical dose for amorphization and TCAD simulation show that all ions are electrically active. Amorphization requires an energy release in the lattice of the order of 25 eV per atom. SRIM simulations show that at 100 keV most of the ions deposit their energy in phonon emission and ionization. Then if we consider the Zn ions depositing all their 100 keV in the silicon substrate, the figures show that implantation doses should be set below $5 \times 10^{13}$ cm$^{-2}$ to avoid any amorphization. With a Zn dose of $5 \times 10^{13}$ cm$^{-2}$ the peak Zn concentration in the $10^{18}$-$10^{19}$ cm$^{-3}$ range. This implies that the post-implantation annealing will be necessary for ion activation only, which is a welcome perspective. Lateral defect control may although be difficult to control.

To make an intermediate conclusion the manufacturing of this pixel device can be consider as feasible, using manufacturing techniques well characterized or implantation methods that should be further studied. High energy Zn, Ge, P implantation with simultaneous thermal anneal, or thermal post-anneal are being planned in Saclay and elsewhere to make a preliminary process for the two TRAMOS options (Deep Impurity and Quantum Box).

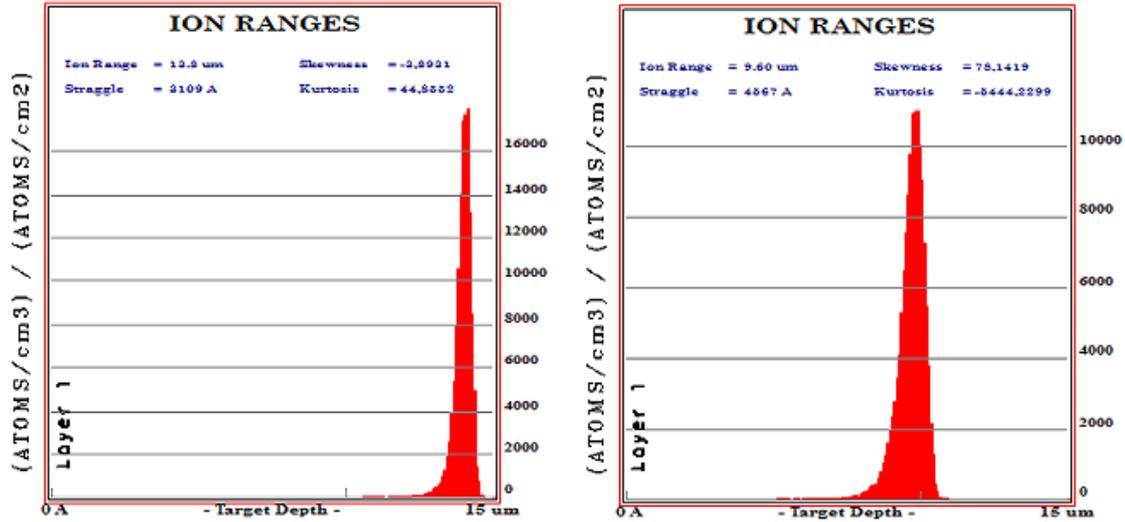

Fig. 8: SRIM simulations for the implantation at high energies (40 MeV) of Phosphorous (left) ions and Arsenic ions (right) into a silicon substrate. The peak introduction rate is approximately 17000 $cm^{-1}$ for P and 11500 $cm^{-1}$ for As. At the same energy the peak of the distribution is located well below 10 μm for P ions whereas for As ions the peak range located approximately at the 10 μm distance from the surface.

## 4. Conclusions

The proposed pixel has been simulated successfully. Its potential characteristics have been evaluated and are better than all present day devices. Different fabrication techniques have been introduced for the fabrication of this DTG pixel detector. A careful analysis has led to the finding of some technological bottlenecks. However they may be overcome with some basic material studies. The physical implementation of the DTG or (Trapping MOS) pixel is within the reach of a present day technologies, despite remaining a challenge with respect to submicron downscaling. The existence of submicron commercially available SiGe processes gives a strong support for this view. This pixel simulated here should be simpler to operate than the former CMOS sensors, because is reduced to one device. Preliminary material and technology work in under way [19]. Its low dimensions combined with the fact that it can operate in a material with high deep level concentrations and a low carrier lifetime strongly favors this pixel design for applications where radiation hardness is sought such as upgrades of the Hadron collider experiments.

## Acknowledgments

The authors are thankful to computer team for providing advice. Many discussions have helped in the advancement of this work specially A. Mesli and G. Regula (IM2NP).